\documentclass[12pt,epsfig]{article}
\date{}
\newcommand{\beq}{\begin{equation}}
\newcommand{\eeq}{\end{equation}}
\newcommand{\beqn}{\begin{eqnarray}}
\newcommand{\eeqn}{\end{eqnarray}}
\newcommand{\dst}{\displaystyle}
\newcommand{\fr}[2]{\frac{{\dst #1}}{{\dst #2}}}

\title{ HIGGS BOSON SEARCH AT PHOTON COLLIDER FOR $M_H=
 140-190$ GEV.} 
\author{I.F.Ginzburg,\ \  I.P.Ivanov\thanks{Novosibirsk State University}\\
Institute of Mathematics, Novosibirsk, e-mail: ginzburg@math.nsc.ru} 

\begin{document}
\maketitle

\begin{abstract}
A Higgs boson within the mass range $M_H=140-190$ GeV can be
discovered at a photon collider in the reaction $\gamma\gamma\to
WW$ (with real or virtual $W$) with the luminosity integral
about 1 fb$^{-1}$. The reasonable resolution in the effective
mass of the $WW$ system is required.
\end{abstract}

{\large\bf Preliminary remarks.}

A discovery of the Higgs boson is one of the main goals for the
next generation of colliders. 

If the Higgs boson mass $M_H$ is larger than $2M_Z$, it can be
discovered at LHC, photon colliders \cite {GKST} or $e^+e^-$
linear colliders \cite{SLACDESY} via the sizable decay mode
$H\to ZZ$. For all types of collisions a background to this
decay mode is rather small. If $M_H<145$ GeV the Higgs boson can
be discovered at $e^+e^-$ linear colliders or photon colliders
via the dominant decay mode $H\to b\bar{b}$ and at LHC -- via
the decay mode $H\to \gamma\gamma$. 

The mass range $M_H=140-190$ GeV is the most difficult one for
the Higgs boson discovery. In this mass range the decay mode
$H\to W^+W^-$ with real or virtual $W$'s ($W^*\to q\bar{q},
e\bar{\nu}, ... $) is dominant, branching ratios of other decay
modes decrease rapidly and their using for the Higgs boson
discovery is very difficult. The use of the $H\to W^+W^-$ decay
at $e^+e^-$ collider is also difficult due to a strong
nonresonant $W^+W^-$ background. (See review \cite{Gun2} and
references therein for more details). As for observation of such
Higgs boson at photon collider, it was noted that "the WW final
state will be a difficult one to use for doing Higgs physics"
\cite{BBC}. The opportunity to see Higgs boson with $M_H \sim
2M_W$ via WW mode at photon collider was noted in
ref.~\cite{Gun1}, where nonrealistic photon spectrum was used
and the interference effects were neglected.
 
In this letter, we show that the Higgs boson with the mass
$M_H=140-190$ GeV can be discovered at the photon collider via
the $H\to W^+W^-$ decay mode. 

To this end we consider both resonant $\gamma\gamma\to H\to W^+W^-$
and nonresonant QED contributions into this process:
\beq
d\sigma\propto  |{\cal M}|^2\equiv |{\cal M}^{0}|^2 + 
2Re({\cal M}^{0*}{\cal M}^H) + |{\cal M}^H|^2.\label{ampl} 
\eeq

Here $|{\cal M}^{0}|^2$ stands for the background QED process
$\gamma\gamma\to W^+W^-$, $|{\cal M}^H|^2$ is the contribution
of $\gamma\gamma\to H\to W^+W^-$ and $2Re({\cal M}^{0*}{\cal
M}^H)$ describes the interference effect of these two mechanism.

The interference contribution is negligible both at high enough
masses of Higgs \cite{GKPS} and far below the nominal threshold
(at $M_{WW}<2M_W$). The interference was neglected in
calculations of \cite{Gun1} for entire mass region. In fact, for
$M_H=200-300$ GeV the interference appeared to be not small
\cite{MTZ}.  Its magnitude increases with decrease in $M_H$.
The range of $M_{WW}$ where the deviation from the
nonresonant background is noticeable, coincides with the Higgs
boson width $\Gamma_H$, therefore this range also decreases with
decrease in $M_H$.  After averaging over a reasonable interval
of $M_{WW}$ the effect was still noticeable providing an
additional way to study the Higgs boson coupling with $W$ bosons
\cite{MTZ}.

We use these considerations as a starting point following a
proposal \cite{Gin96}.\\

{\large\bf\boldmath The WW production cross section at
$140<M_{WW}<190$ GeV.} 

We consider initial photon state with total helicity zero
(helicities of colliding photons $\lambda_1 = \lambda_2 =\pm
1$).  Only a photon state with such quantum numbers can produce
the Higgs boson.  Helicity amplitudes for the nonresonant
process (in the Born approximation) for such initial state are
presented in ref.~\cite{BB}.  In this case the helicities
$\lambda $ of produced $W$ bosons coincide. One has (the
subscript corresponds to helicity of produced $W$, $\Lambda$ is
the product of the photon and $W$ helicities):
\beq 
{\cal M}^0_0 =-\fr{2\pi\alpha sx}{p_\bot^2 +M_W^2};\quad
{\cal M}^0_{\pm 1} = \fr{ \pi\alpha s}{p_\bot^2 +M_W^2}
\left(2\Lambda\sqrt{1-x} +2-x\right)\quad
\left(x=\fr{4M_W^2}{s}\right).\label{QEDamp}
\eeq
The amplitude ${\cal M}^H$ is written in the standard form via
the well known amplitude of Higgs boson two photon decay
\cite{vain}. 

In practice we deal with the production of four fermions in the
final state produced via intermediate W boson state. The
distribution in invariant masses for such final state has a
maximum at $M_W$ with the width $\Gamma_W$ and long tails away
from the maximum. It can be described by the well known
Breit--Wigner spectral density for the W boson.  One can view
this as the production of $W^*$ --- a virtual $W$ boson with the
mass not equal to $M_W$. Because of this, the cross section does
not vanish below the nominal threshold $M_{WW}^{thr}=2M_W$. The
most important effect comes from the corresponding modification
in the final phase space \cite{GKPS}. It can be described via
replacement of the relative velocity of $W$ bosons in the center
of mass frame
$$
\beta=\frac {1}{s}\sqrt{(s-s_1-s_2)^2- 4s_1s_2}
$$
by its convolution with above spectral density for the $W$
boson $\varrho(M^2)$:
$$ 
\beta\to\tilde {\beta} = \int
ds_1ds_2~\varrho(s_1)~\varrho(s_2)~\beta(s,s_1,s_2)~\theta
(s_1)~\theta (s_2)~\theta (\sqrt {s} -\sqrt{s_1}-\sqrt{s_2}).
$$

Close to the threshold the $\gamma \gamma \to W^+ W^-$ cross
section for stable $W$ bosons can be approximated as:
$$
\sigma \propto |{\cal M}|^2_{s=4M_W^2}~\beta, 
$$ 
We do not have explicit formulas for the background process
below the threshold. Therefore, we restrict ourselves to the
approximation where helicity amplitudes are calculated at the
nominal threshold $2M_W$ neglecting the effect of the finite
width. Finally, we approximate cross section near the threshold 
by equation
$$
\sigma \propto |{\cal M}|^2 _{s=4M_W^2}~\tilde {\beta},
$$ 

The accuracy of this approximation can be checked by comparing
exact calculation for the Higgs boson width $\Gamma_{H\to
W^+W^-}^{prec}$ (based on ref.~\cite{Hwid}) with our
approximation. Let us denote the Higgs boson width obtained in
the framework of our approximation as $\Gamma_{H\to W^+W^-}^
{phas}$.  We find then that the ratio $\Gamma_{H\to
W^+W^-}^{prec}/\Gamma_{H\to W^+W^-}^{phas}$ varies from 1.2 for
$M_H=160$ GeV to 1.6 for $M_H=140$ GeV. Note, that the Higgs
boson width itself decreases by a factor of $20$ for $M_H = 140$
GeV in comparison with its value at $M_H=160$ GeV, so one can
conclude that our approximation works well.

Let us now describe the results of the calculation. In our
estimates we neglected angular dependence of the cross sections
(since we are working in the threshold region) and integrated
over production angle in the center of mass frame within the
range $20^{\circ} <\theta <160^{\circ}$. This represents $94 \%$
of the total solid angle.

In Fig. 1 we show $\gamma \gamma \to W^+ W^-$ cross section for
various $M_H$ as a function of the invariant mass $M_{WW}$ of
the $W^+W^-$ pair for the initial state with total helicity
zero.  For $M_H\geq 180$ GeV the two--peak behavior observed in
\cite{MTZ} is seen clearly.  The interference term in eq.
(\ref{ampl}) is large for such Higgs boson masses. The height of
the peak grows with decrease in $M_H$ but its width
($\approx\Gamma_H$) becomes smaller and smaller.

With subsequent decrease in $M_H$, the background QED process is
suppressed by the phase space and the interference also becomes
small in comparison with the Higgs boson contribution.\\

{\large\bf The ``experimental'' cross sections.}

After averaging over initial energy distribution of the photons,
the Higgs boson signal decreases and seems to be hardly
observable. To circumvent this problem, the observation of $W$
bosons in the final state is mandatory (perhaps, via quark
jets). In order to account for a finite resolution in $M_{WW}$,
we consider a ``smeared'' cross section, similar to \cite{MTZ}: 
\beqn
\fr{d\sigma}{dM_{WW}^{meas}} = \int \fr{dM_{WW}}{\sqrt{2\pi}\ d} 
\exp \left[-\fr{(M_{WW}^{meas}-M_{WW})^2}{2d^2}\right] 
\fr{d\sigma}{dM_{WW}}\label{sm} 
\eeqn 
The results for $d=5$ GeV are presented below. 

Let us stress that we do not perform any convolution with
initial photon spectra. It is because of the fact that the real
form of these effective spectra strongly depends on the details
of the conversion design. Note also, that it can not be obtained
by a simple convolution of individual spectra \cite{GKST}. Real
energy distribution of colliding photons should be measured for
every energy of collider\footnote{ The effect of averaging over
initial photon spectrum without reconstruction of $M_{WW}$ can
be simulated by using $d\approx 15$ GeV in eq.~(\ref{sm}) (cf.
\cite{GKST},\cite{SLACDESY}).}. We have in mind ``monochromatic''
variant of the photon collider with the effective width of the
$\gamma\gamma$ energy distribution $\sim 15$~\% \cite{GKST,SLACDESY}. 

The results of this calculation are shown in Fig. 2 (above the
nominal threshold), in Fig. 3 (below the nominal threshold)
and in the Table.

In view of a very high degree of photon polarization expected at
photon colliders, we present the Figures for completely
polarized photon beams: $<\lambda_1> = <\lambda_2>=\pm 1$.
Higgs boson signal deteriorates when $<\lambda_1><\lambda_2>$
decreases. Some result for $<\lambda_i> \ne 1$ are presented in
the Table for ``realistic'' mean helicities of the photons
\cite{GKST,SLACDESY}.
\begin{table}[tb]
\begin{center}
\begin{tabular}{|c|c|c|c|c|c|}  \hline
Higgs boson mass& Characteristic &&&&\\
 $M_H$, GeV & $M_{WW}$, GeV & $<\lambda>$ & 
     $\sigma^{bkgd},pb$ & $\sigma^{tot},pb$ & 
     $\fr{\sigma^{tot}-\sigma^{bkgd}}{\sigma^{bkgd}}$ \\  \hline \hline
 140 & 140 & 1.0 & 0.11 & 0.98 & 8.0 \\
     &     & 0.9 & 0.17 & 0.96 & 4.6 \\ \hline
 150 & 150 & 1.0 & 0.42 &2.10 & 4.0 \\ 
     &     & 0.9 & 0.60 & 2.13 & 2.6\\ \hline
 160 & 159.7 & 1.0 & 3.65 & 6.81 & 87\% \\
     &     & 0.9 & 4.86 & 7.74 & 59\% \\ \hline
 170 & 169.3 & 1.0 & 12.6 & 15.5 & 23\% \\
     &     & 0.9 & 16.0 & 18.5 & 16\% \\ \hline
 180 & 178.5 & 1.0 & 21.8 & 23.7 & 8.7\% \\
     &     & 0.9 & 25.3 & 27.0 & 6.7\% \\ \hline
 190 & 185 & 1.0 & 26.0 & 27.0 & 3.8\% \\
     &     & 0.9 & 29.9 & 30.8 & 3.0\% \\
     & 200 & 1.0 & 37.3 & 36.5 & -2.1\% \\
     &     & 0.9 & 40.2 & 39.4 & -2.0\% \\ \hline 
\end{tabular}
\caption{ $\gamma\gamma\to W^+W^-$ cross sections for
various $M_H$ and $<\lambda>$} 
\end{center}
\end{table}

Above the nominal threshold the curves in Fig. 2 exhibit a
characteristic shape. A deviation from the QED background is
very sensitive to the presence of the Higgs boson. Therefore,
such behavior of the experimental curve can be considered as an
evidence for the existence of the Higgs boson in this mass
region. Here the background QED cross section varies from 3.6 pb
for $M_{WW}=160$ GeV to 30 pb for $M_{WW}=190$ GeV. The Higgs
boson production cross section adds another 3 pb for $M_H=160$
GeV (87\%). For $M_H=190$ GeV an interference term in eq. (\ref
{ampl}) delivers 1.8 pb (6\%) to the total cross section.

Below the threshold Higgs boson is observed even more clearly
(Fig. 3). The total cross section is about 1 pb in the whole
region. It is larger than the cross section for the background
process by almost a factor of 2 for $M_H=160$ GeV and by a
factor of 9 for $M_H=140$ GeV. This result is in qualitative
agreement with that in ref.~\cite{Gun1}.

This behavior is not surprising. The total Higgs boson
production cross section averaged over the range of $M_{WW}$
larger than the total width of the Higgs boson is equal to
$\sigma_{\gamma \gamma\to H} = 4\pi^2\Gamma_{\gamma\gamma}/
M_H^3$. The Higgs boson width drops out in this result. Roughly
speaking, the cross section is almost independent of $ M_H$ in
the considered mass range.  At the same time, down to $M_H =
145$ GeV the $W^+W^-$ decay mode still remains dominant. On the
other hand, the QED background decreases fast with decrease in
$M_{WW}$ in the considered mass range. For this reason the two
peak behavior, observed for $M_H>180$ GeV, changes to a single
resonance peak near and below the threshold and the Higgs boson
signal becomes dominant in this region. 

As was discussed above, more precise calculation of the Higgs
boson width would enhance the effect by a factor $\sim 1.6$ for
$M_H=140$ GeV.  We expect that the signal-to-background ratio
will be close to the one calculated in this paper.\\

{\large\bf Discussion. Final remarks.}
 
We conclude that the Higgs boson with the mass $M_H=140-190$ GeV
can be observed and studied in the process $\gamma \gamma \to
W^+ W^-$ at future $\gamma \gamma$ colliders by measuring the
invariant mass of produced $W^+W^-$ system with reasonable
resolution. A luminosity integral required for this observation
is $\sim 1$~fb$^{-1}$. This value seems to be sufficient even
accounting for corrections due to detection efficiency etc., and
the electromagnetic 4 jet production\footnote{ The cross section
is saturated by events with effective mass of decay products of
one $W$ boson lying within the resonance peak. Above the
threshold the same is valid for the other $W$ boson, below the
threshold the corresponding effective mass is below the
resonance peak.}.

Moreover, we found that for $M_H\leq 160$ GeV the effect of the
Higgs boson is still seen after being smeared over $15$ GeV.
This means that the signal of the Higgs boson lighter than $160$
GeV can be observed even in the total $\gamma\gamma\to W^+W^-$
cross section without restrictions on the $W^+W^-$ invariant
mass.

Additionally, if the Higgs boson mass is near 140 GeV, one can
hope to observe Higgs boson decays into $b\bar b$ and $W^+W^-$.
It provides an opportunity to compare the Higgs boson coupling
to quarks and to gauge bosons and in this way to test the Higgs
boson origin of particle masses.

Last, similar calculation can be easily repeated for $\gamma
\gamma \to H\to ZZ$ process below its nominal threshold
(the nonresonant background is negligible here). The
corresponding cross sections are $\sim 50$ fb. Therefore,
provided the luminosity integral $\sim$ 10 fb$^{-1}$, one can
also observe a Higgs boson in ZZ decay mode in the whole region
of $M_H$ considered here. The comparison of Higgs boson
couplings with different gauge bosons will be essential test of
SM concerning Higgs mechanism of the mass origin.

Our analyses at this point should be considered as an estimate
of the effect. We are sure that our estimates give correct order
of magnitude prediction for the signal to background ratio.
However, more studies are required to arrive at absolute
predictions. At the first stage of such studies Born
approximations for all quantities should be improved by taking
into account relatively large corrections due to the finite $W$
width in $\gamma\gamma\to H$ amplitude \cite{MelYak}.\\

This work is supported by grants INTAS -- 93--1180~ext and RFBR
-- 96--02--19114. We are grateful to A.E.~Bondar, J.F.~Gunion,
K.V.~Melnikov, V.G.~Serbo and V.I.~Telnov for discussions.

\end{document}